\documentclass[conference]{IEEEtran}
\usepackage[noadjust]{cite}

\ifCLASSINFOpdf
  \usepackage[pdftex]{graphicx}
\else
\fi
\usepackage[cmex10]{amsmath}
\begin{document}
%
\title{Modelling Stock-market Investors as \\Reinforcement Learning Agents [Correction]}

\author{\IEEEauthorblockN{Alvin Pastore}
\IEEEauthorblockA{Department of Computer Science\\
University of Sheffield\\
Sheffield, United Kingdom\\
Email: apastore1@sheffield.ac.uk}
\and
\IEEEauthorblockN{Umberto Esposito}
\IEEEauthorblockA{Department of Computer Science\\
University of Sheffield\\
Sheffield, United Kingdom\\
Email: acp12ue@sheffield.ac.uk}
\and
\IEEEauthorblockN{Eleni Vasilaki}
\IEEEauthorblockA{Department of Computer Science\\
University of Sheffield\\
Sheffield, United Kingdom\\
Email: e.vasilaki@sheffield.ac.uk}
}


%


\maketitle

\begin{abstract}
Decision making in uncertain and risky environments is a prominent area of research. 
Standard economic theories fail to fully explain human behaviour, while a potentially promising alternative may lie in the direction of Reinforcement Learning (RL) theory.
We analyse data for 46 players extracted from a financial market online game and test whether Reinforcement Learning (Q-Learning) could capture these players behaviour using a risk measure based on financial modeling.  
Moreover we test an earlier hypothesis that players are ``na{\"i}ve'' (short-sighted). 
Our results indicate that a simple Reinforcement Learning model which considers only the selling component of the task captures the decision-making process for a subset of players but this is not sufficient to draw any conclusion on the population. 
We also find that there is not a significant improvement of fitting of the players when using a full RL model against a myopic version, where only immediate reward is valued by the players. This indicates that players, if using a Reinforcement Learning approach, do so na{\"i}vely.
\end{abstract}



%
\IEEEpeerreviewmaketitle

\section{Introduction}

\begin{figure*}[!ht]
\centering
\includegraphics[width=7.25in]{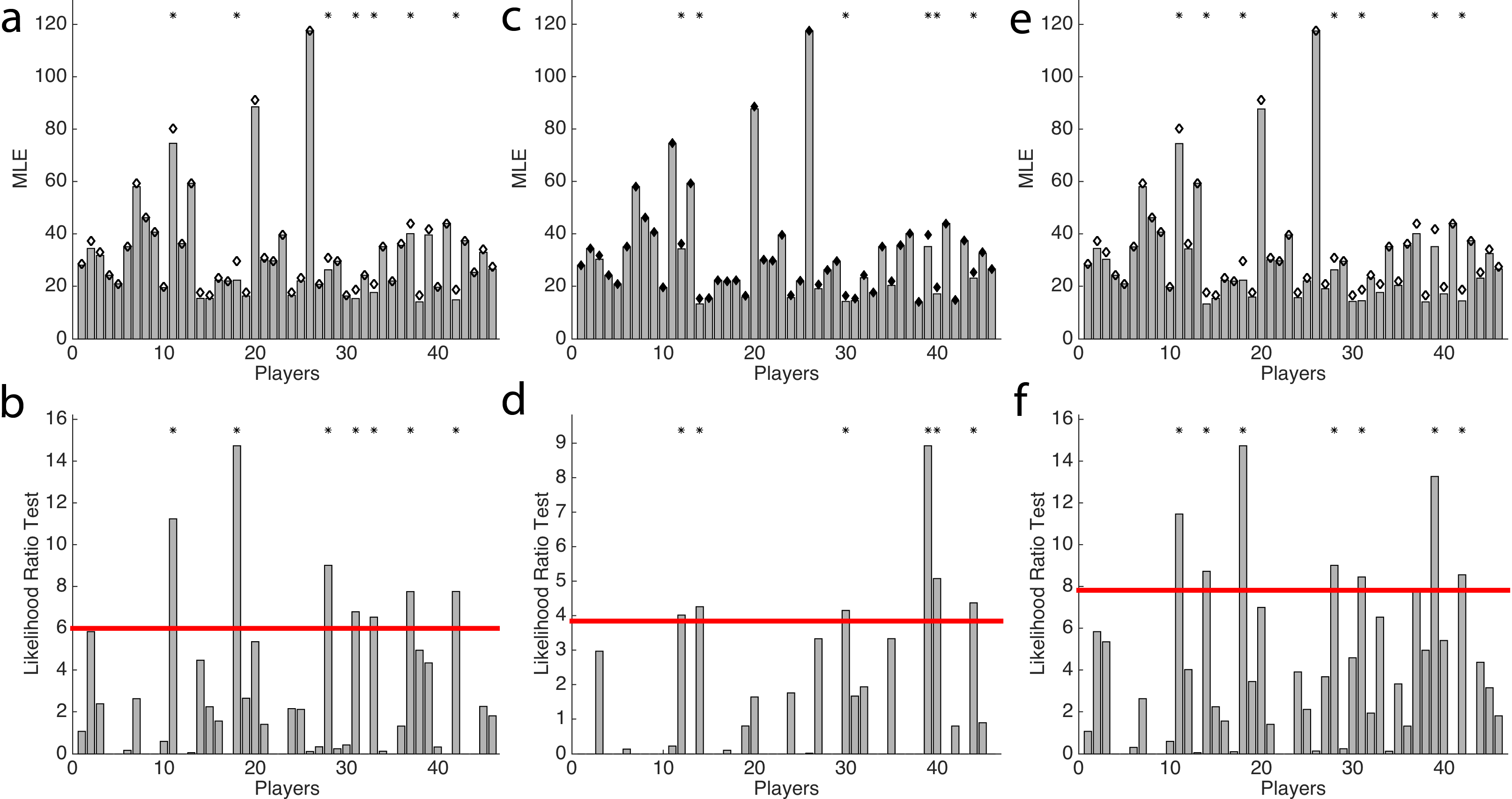}
\caption{Models comparisons. Figure (a) shows the goodness of fit (Maximum Likelihood Estimate) of the myopic model (2 free parameters, with $\gamma = 0$) against the random model. Each bar represents the corresponding player's MLE while the hollow diamond represents the MLE of the random model. 
The height of the bars differs greatly as the players have different amount of transactions. 
The asterisks indicate the players better fitted by the myopic version of our RL model as opposed to a random model. 
For 15$\%$ of the players the myopic model fits better than random (players 11, 18, 28, 31, 33, 37, 42 with p-value $<$ 0.05 calculated with the Likelihood ratio $\chi^2$ test at 95$\%$ confidence level). 
Figure (b) shows the same comparison, this time with the result of the Likelihood Ratio Test portrayed on the y-axis. The horizontal line represent the threshold (critical value) for statistical significance which depends on the degrees of freedom. 
For myopic model versus random $dof = 2$ and critical value is 5.991.
Figures (c) and (d) show the comparison for the fitting of the full RL (3 free parameters) model against the simpler version ($\gamma = 0$) model. 
In (c) each bar represent a player's full RL model MLE while the filled diamond represents the MLE of the myopic model. Players who are better fitted by a full model are marked with an asterisk (statistically significant at 95$\%$ confidence with the same test as the previous comparison). Even if there is a statistical improvement in the fitting for 6 players (12, 14, 30, 39, 40, 44) this does not translate in a better overall performance of the full RL model when it is compared to the random model. Figures (e) and (f) show this comparison. The full RL fits better than random for 7 players (11, 14, 18, 28, 31, 39, 42). Some of these are already captured by the simpler model, only two (14 and 39) were not already captured by the myopic model. At the same time the full model loses two players (33 and 37). This is due to the fact that this comparison has a higher critical value (7.815 with $dof = 3$). These results have been tested for the entire population using Binomial Proportion Confidence Intervals calculated with Clopper-Pearson method, the results are negative as only 7 out of 46 players are better than random (Fig. \ref{fig:population_statistic} in the Appendix)}
\label{fig:models_comparisons}
\end{figure*}

One of the most challenging fields of research is human decision-making. 
Understanding the processes involved and trying to predict or replicate behaviours has been, historically, the ultimate goal of many disciplines. 
Economics for example, has a long tradition of trying to formalise human behaviour into descriptive or normative models. 
These models have been employed for several years (e.g. Expected Utility model \cite{J.vonNeumann1947}) but have been proven to be inadequate \cite{Starmer2000,ShaneFriederickGeorgeLoewenstein2002,Tversky1974,Kahneman1991}, giving rise to new research areas like behavioural and experimental economics. 
Psychology as well, is natively concerned with decision-making. 
Sequential decision problems have been used to evaluate people's  risk attitude, in order to predict actual risk proneness in real life scenarios \cite{Hoffrage2003,Pleskac2008,Wallsten2005}. 
While economics and psychology are focused on the high-level manifestations and implications of decision-making, neuroscience aims at understanding the biological machinery and the neural processes behind human (or animal) behaviour \cite{Britten1996,Britten1992,Gold2001,Gold2002,Gold2007,Shadlen1996,Shadlen1996a}.

Recently these fields of research have started to collaborate, contributing to the rise of an emerging multi-disciplinary field called neuroeconomics \cite{Sanfey2006, Glimcher2004, Loewenstein2008, Barraclough2004, Soltani2006}.
This discipline approaches the problem from several perspectives and on different levels of abstraction. 
RL is a theoretical framework \cite{Sutton1998}, extensively used in neuroeconomics literature for addressing a wide array of problems involving learning in partially observable environments \cite{Doya2007, Tesauro1994, AbbeelPCoatesAMorganQNg2006, Hafner2007, Walker2000, Mnih2015}. 
RL is based on the concept of reward/punishment for the actions taken. 
The agents act in an environment of which they possess only partial knowledge. 
To be able to achieve the best behaviour, i.e. maximise their reward, the agents have to learn through experience and update their beliefs. 
Learning happens as a result of the agent's interpretation of the interactions with its surroundings and the consequences of a ``reward'' feedback signal.
The ability of this framework to model and therefore understand behavioural data and its underlying neural implications, is of pivotal importance in decision making \cite{Dayan2008}. 

RL can accurately capture human and animal learning patterns and has been proven effective at describing the functioning of some areas of the human brain, like the basal ganglia, and the functions of neurotransmitters such as dopamine \cite{Schultz1997, Doya2000, Daw2006, Doya2007}.
One of the most remarkable similarities between biological functioning and RL models is the one about Temporal Difference (TD) error \cite{Sutton1998, Watkins1989, Werbos1990, Hikosaka2006} and the activation of mid-brain dopamine neurons \cite{SchultzWRomoRLjungbergTMirenowiczJHollermanJRandDickson1995,Suri1998, Waelti2001, Satoh2003, Nakahara2004, Morris2006}. 
These findings supported the notion that TD Learning is implemented in the brain with dopaminergic neurons in the striatum \cite{Schultz1997, Barto1995, Montague1996, Samejima2005, Kawagoe1998, Kawagoe2004, Hikosaka2006, Schultz2002, Day2007, Costa2007, Hyman2006, Joel2002, Wickens2007}, making it a reasonable first choice for a modelling attempt.
Humans and animals are very advanced signal detectors whose behaviour is susceptible to changes in the rewards resulting from their choices \cite{Stocker2006,Kording2007}. 
Both neuroscience and psychology have extensively employed tasks in which the exploration-exploitation trade-off was of crucial importance \cite{Daw2006a, Soltani2006,Luksys2009,Frey2015, Benartzi2015}. 
It is crucial for the individuals to maximise their reward using the information at their disposal but to do so advantageously they need to learn which actions lead to better rewards.
Decision making in uncertain environments is a challenging problem because of the competing importance of the two strategies: exploitation is, of course, the best course of action, but only when enough knowledge about the quality of the actions is available, while exploration increases the knowledge about the environment.

A complicated task that encompasses all these features is stocks selection in financial markets, where investors have to choose among hundreds of possible securities to create their portfolio. 
Stock trends are non-monotonic because they are not guaranteed to achieve a global maximum and the future distribution of reward is intrinsically stochastic. 
After purchasing a stock, investors are faced with the decisions on when to sell it (Market timing problem \cite{Benartzi2015}). 
To be able to achieve the best return from their investments, people need to be careful in  considering how to maximise their profit in the long term and not only in a single episode. 
We speculate that RL is part of the decision making process for investors. 
This speculation is supported by Choi et al. \cite{Choi2009}, who studied individual investors decisions on 401(k) savings plans.
Over the years, investors could decide to increase or decrease the percentage of their salary to commit to this retirement plan.
Their results suggest that investors' decisions are influenced by personal experiences:
they show that those investors who have experienced a positive reward from previous investment in their 401(k) fund, tend to increase the investment in that particular product, compared to those who experienced a lower reward outcome. 
This kind of behaviour follows a ``na{\"i}ve reinforcement learning'' and is in contrast with the disposition effect \cite{Barber2013, Odean1998}(the unwillingness of investors to sell ``losing'' investments).
Huang et al. investigated how personal experience in investments affects future decisions about the selection of stocks \cite{Huang2012}. They used data that spans from 1991 to 1996, from a large discount broker. 
Again, the pattern of repurchasing financial product which yielded positive return was found. 
As Huang suggests, by understanding the way past experience affects investors' decisions, it might be possible to make predictions about the financial markets involved.
RL has also been used, with promising results, to develop Stock Market Trading Systems \cite{Chen2007, Lee2001, O2006, Moody2001} and to build Agent Based Stock Market Simulations \cite{Rutkauskas2009}. 
While these works use RL to predict future prices, they do not try to describe human behaviour. 
With these notions as background we decided to investigate and try to model human choices in a stochastic, non-stationary environment.
We hypothesise that RL is a component of decision making and to test this we compare two RL models against a purely random one. Our modelling attempts are based on two assumptions. First, we assume that risk is a proxy of the internal representation of the actions for some players. To test this we use a measure of systematic risk widely used in finance and economics to categorise the different choices into three discrete classes. 
We also assume that the reward signal is based on the cash income arising from the sales an investor makes. 
This assumption follows a widely researched behaviour referred to as ``disposition effect'' in literature \cite{Barber2013, Odean1998}, the tendency of individual investors to sell stocks which increased in value since when they were purchased, while holding onto the stocks which lost value. 
This phenomenon is stronger for individual investors but it also exhibited by institutional investors, such as mutual funds and corporations \cite{Brown2006,Barber2007,Frazzini2006,Grinblatt2001,Heath1999,Dean1998,Shapira2001}.
Following these indications we mapped the sell transactions to a reward signal to fit our models.
Finally, we hypothesise that not all players are short-sighted, to test this we compare a full RL model (3 free parameters) against a myopic model (2 free parameters, no gamma). The difference is that the latter can be considered a na{\"i}ve RL as it does not take into account future rewards, it only seeks to maximise immediate rewards. 

\begin{figure}[t]
\includegraphics[width=3.5in]{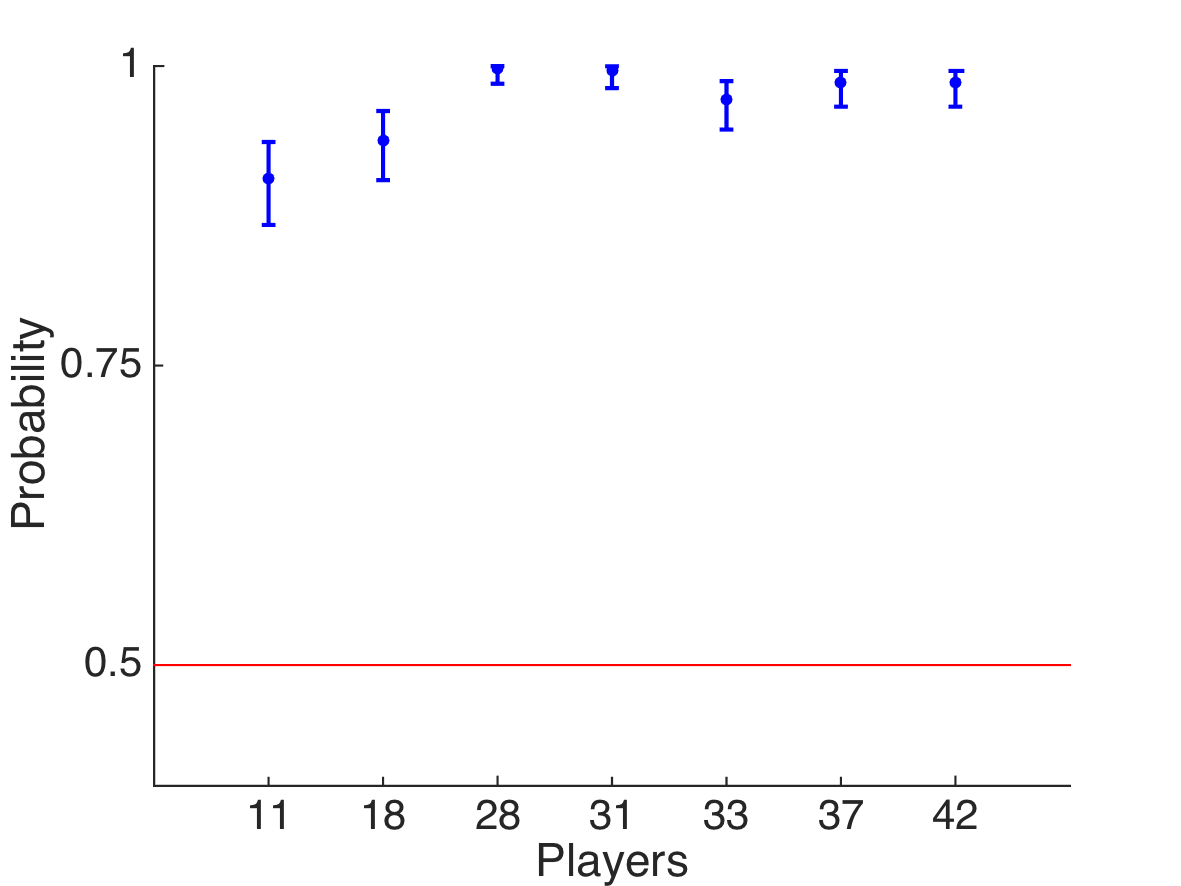}
\caption{Comparison of the ranked vs scrambled discretisation. Test for the assumption that investors use risk to internally classify the different stocks in discrete degrees of risk. The errorbars represent the probability and 99$\%$  confidence intervals of the comparison of the stock discretisation based on the $\beta_F$ risk measure as opposed to 500 randomly generated scrambled discretisations.
The 7 players for which the RL model outperformed the random model all have a statistically significative probability of using a representation of actions discretisation based on risk. 
The comparison of the entire dataset resulted in 31 players out of 46 ($\sim 67 \%$) to have a probability and CI above chance.
}
\label{fig:ranked_v_scrambled}
\end{figure}

\section{Method}
\subsection{Dataset}
The dataset has been extracted from the publicly accessible online trading simulation game VirtualTrader\footnote{http://www.virtualtrader.co.uk - Copyright IEX Media Group BV}, which is managed by IEX Media Group BV in the Netherlands. 
Players can subscribe for free and start playing the game with an assigned virtual cash budget of 100k GBP. 
The players will then pick the stocks they prefer from the FTSE100 stock index pool (107 stocks at the time of data collection) and create their own portfolio. 
These competitors are ranked according to the return of their investment. This is composed of ``holdings'' and ``cash''. The former represent the shares possessed by a player while the latter is the amount of money not invested (i.e. deriving from sold stocks or never invested).
The simulation follows real world data evolution, for example price fluctuations and price splits. 
The delay is usually in the order of 10-15 minutes and the player can access a visual representation of the stocks time series. 
All the transactions are stored for each player.
For this study we considered transactions that span from the 1st of January 2014 to the 31st of May 2014. 
This time period has been chosen because at that time the player ranking was determining the winner of the monthly prize giveaway.
Two possible rewards can be identified: a psychological one, consisting of the ranking position and a tangible one being the prize for the highest achiever.

The transactions have been stored in a database in order to be manipulated and used to fit models with different combination of free parameters.
The rows are structured in 6 fields: Date, Type, Stock, Volume, Price and Total.
The dataset initially contained about 100k transactions that were reduced to about 1.4k.
This was due to preprocessing, which removed the many instances of inactive players who played only at the beginning and/or at the end of the time frame considered.
In the final version of the dataset there are 46 players.
The average amount of transactions per player is 30. 
The player who played the most during the six months performed 107 transactions.
We considered the full amount of transactions each player operated in the game.

\subsection{Reinforcement Learning Setup}
We adopted a widely used off-policy RL framework called Q-learning \cite{Sutton1998}.
The learning rule of this model is:

\begin{equation}\label{q-learning_equation}
	\Delta Q(s_t,a_t) =  \alpha \bigg [ r_{t+1} + \gamma  \max\limits_{a} Q(s_{t+1},a) - Q(s_t,a_t) \bigg]
\end{equation}

where $Q(s_t,a_t)$ represents the value of action $a$ while in state $s$, at time $t$.
$\alpha \in [0,2]$ is the step-size parameter and controls the rate of learning. 
$\gamma \in [0,1]$ is the discount factor and represents how far-sighted the model is,
It encodes how much a future reward is worth at time $t$.  When $\gamma = 0$ only immediate rewards are taken into account by the player.

To test this framework the task has been mapped as follows. 
There are two states (win, loss) calculated according to the profit of the player (details in equations \ref{eq:profit_equation} and \ref{eq:State_equation}). These two states reflect the dichotomy rooted in the Prospect Theory's value function gain/loss spectrum \cite{Kahneman1979}.

Since all players begin with the same initial budget our calculation of the profit uses the returns accumulated by selling stocks. 
This choice reduces the scope of the model, focusing on the cash component of the players assets. This will be referred to as the ``Sell'' model. 
The actions are mapped to the stocks available for trading. 

In order to avoid dimensionality issues, 107 stocks for 2 states give rise to 214 potential actions, we decided to classify the stocks in 3 classes of risk using a widely used financial modelling measure, $CAPM \  Beta$. 
The acronym stands for Capital Asset Pricing Model, a model developed by Sharpe\cite{Sharpe1964} used to explain the relationship between the expected return of a security and its risk .
In this report we will refer to financial volatility measure $CAPM \  Beta$ as  $\beta_F$:

\begin{equation}\label{beta_equation_covariance_variance}
	\beta_F = \frac{Cov(r_a,r_b)}{Var(r_b)}
\end{equation}

This financial modelling measure quantifies the volatility of a security in comparison to the market or a reference benchmark \cite{Beninga2000}. 
Relatively safe investments like utilities stocks (e.g. gas and power) have a low $\beta_F$, while high-tech stocks (e.g. Nasdaq or MSH Morgan Stanley High-Tech) have a high $\beta_F$.

As an example, the $\beta_F$ of the index of reference (that represents the portion of the market considered) is exactly 1. 
A $\beta_F \in (0,1)$  indicates that the asset has a lower volatility compared to the market or low correlation of the asset price movements compared to the market. 
While if  $\beta_F > 1$ it signifies an investment with higher volatility compared to the benchmark. 
Following the previous example, high-tech securities with a $\beta_F > 1$ could yield better returns compared to their benchmark index, when the market is going up. 
This also poses more risk because in case the market loses value, the security would lose value at a higher rate than the index. 

$\beta_F$ is considered a measure of the systematic risk and can be estimated by regression.
Considering the returns of an asset $a$ and the returns of the corresponding benchmark $b$:

\begin{equation}\label{beta_equation_regression}
	r_a = \alpha + \beta r_b
\end{equation}

$\beta_F$ has been calculated for each stock in the FTSE100 at the time of the game by considering daily returns in the year between 1st June 2013 and 31st May 2014.
The measure associated with each stock is used to rank them and subdivide them in three classes, containing respectively 36, 36 and 35 stocks each. 

Reward $r$ at time step $t+1$ is defined by the gain (or loss) made in a sell transaction. 
Buying transactions are kept into account to track players portfolios and to calculate the price difference. They were not used as actions, but we might extend our modelling scenario by integrating a ``Buy'' model in the future and consider purchase actions by changing the reward scheme.
The reward is calculated as:

\begin{equation}\label{eq:reward_calculation}
	r_{t+1} = v_{t+1}\left (p_{t+1} - \frac{1}{t} \sum_{i=1}^{ t} v_i  p_i   \right )
\end{equation}

where $v_{i}$ and $p_{i}$ are the volume and the price of the stock traded at the i-th time step. 
The second term of the difference is a weighted average of the stock prices at previous times. 

To avoid numerical instabilities, reward has been flattened with a sigmoid function into the range $[-1,+1]$.
Specifically a hyperbolic tangent has been used, with $\rho  = 500$ to capture most of the variability of the rewards, only flattening the extreme values. This choice is in line with prospect theory value function which is concave for gains and convex for losses \cite{Kahneman1979}.

\begin{equation}\label{eq:hyperbolic_tangent}
   tanh(r) = \frac{1-e^{-r/\rho }}{1+e^{-r/\rho }}
\end{equation}

As in this study we focused on the sell subset of the players interactions, the states are based solely on profit, which in turn is based on the reward of the sell transactions. The profit and states are defined as:

\begin{equation}\label{eq:profit_equation}
	Profit = \sum_{t} tanh(r_t)
\end{equation}
	
\begin{equation}\label{eq:State_equation}
	State=\begin{cases}
    0, & \text{if $Profit<0$}.\\
    1, & \text{otherwise}.
  \end{cases}
\end{equation}

The RL framework is composed of a learning model (eq. \ref{q-learning_equation}) and an action-selection model that is responsible for picking the best action.
In our setup the former is Q-learning and the latter is Soft-Max.
An action $a$ is picked at time $t$ with probability:

\begin{equation}\label{eq:softmax_equation}
	P(a_t) = \frac{e^{Q_{t} (a) \beta}}{\sum_{b=1}^{n} e^{Q_{t}(b) \beta}}
\end{equation}

where $n$ is the number of available actions (i.e. 3 in this study) and $\beta$ is the inverse temperature parameter and represents the greediness of the action-selection model. In the limit $\beta \rightarrow 0$ the actions become equiprobable and the model reverts to random. 
Higher values of $\beta$ approximate a greedy model which picks the best known action (fully exploitative).
The full model has 3 free parameters: $\alpha$ (step-size parameter or learning rate), $\beta$ (exploration-exploitation trade-off) and $\gamma$ (discount factor). 
For this study we used a bounded gradient descent search with 27 combinations of initial guess points. These are the combination of values of the free parameters from where the search starts. By having different entry points we hope to reduce the chance of the search getting stuck in a local minimum solution. The search has been performed with the following boundaries:

\begin{itemize}
\item $\alpha \in (0.0001, 2)$ 
\item $\beta \in (0, 50)$ 
\item $\gamma \in (0, 0.9999)$ (for the myopic model $\gamma = 0$)
\end{itemize} 

The entry points are the combinations rising from the following values:

\begin{itemize}
\item $\alpha \in  \left \{0, 0.5, 2\right \}$
\item $\beta \in  \left \{0, 25, 50\right \}$ 
\item $\gamma \in  \left \{0, 0.5, 0.9999\right \}$ 
\end{itemize} 

The search results have been obtained on \texttt{python 2.7.9} and \texttt{scipy.optimize.minimize} with \texttt{scipy 0.17.1}.

\subsection{Model Testing Routine}

Maximum Likelihood Estimate has been used as a measure of the model fitness, following Daw's comprehensive analysis of methodology \cite{Daw2009}.
MLE is the appropriate method to assess model performance because it evaluates which set of model parameters are more likely to generate the data using a probabilistic approach. 
Data likelihood is a powerful method because it keeps into account the presence of noise in the choices. It does so by using probability estimates for the potential actions. 

Given a model $M$ and its corresponding set of parameters $\theta_M$ the likelihood function is defined as $P(D|M,\theta_M)$, where $D$ is the dataset (the list of choices and the associated rewards).
Applying Bayes' rule:

\begin{equation}\label{eq:bayes_rule_likelihood}
	P(\theta_M|D,M) \propto P(D|M,\theta_M) \cdot P(\theta_M|M)
\end{equation}

The left hand side of the proportionality is the posterior probability distribution over the free parameters, given the data. This quantity is proportional to the product of the likelihood of the data, given the parameters and the prior probability of the parameters. Treating the latter as flat we obtain that the most probable value for $\theta_M$ (the best set of free parameters) is the Maximum Likelihood Estimate (MLE), that is the set of parameter which maximises the likelihood function, $P(D|M,\theta_M)$ and it is commonly denoted $\hat\theta_M$. The likelihood function is maximised through the following process. 
At each timestep, for every action, the observation model (Soft-Max) estimates a probability. These probabilities are then multiplied together. To avoid numerical problems that could arise when multiplying probabilities, the sum of their logarithm is calculated instead. 
The negative of this value, also known as Negative Log-Likelihood, is then used. The aim is then to minimise this quantity, which is the equivalent of maximising the likelihood function, $P(D|M,\theta_M)$. In the future we will refer to the Negative Log-Likelihood as MLE for simplicity, keeping in mind that lower values represent better fit.

The values of MLE generated represent the goodness of fit of the model with its associated set of parameters. To compare the selected model with a random model and for statistical significance we adopted the Likelihood Ratio Test \cite{Crandall1997}. This statistical test uses the likelihood ratio to compare two nested models and takes into account the different number of free parameters of the two. It encapsulates this information, when testing for significance, using the difference of the two amounts as degrees of freedom for the chi-square ($\chi^2$) test. Since the test statistic is distributed $\chi^2$ it is straightforward to estimate the p-value associated with the $\chi^2$ value.

The baseline for comparison is a random model which has 0 parameters as there is no learning involved and the action-selection policy is random ($\frac{1}{3}$ chance of picking any of the three stock bins). The first comparison is between this random model and the simpler of the proposed models, which has only two free parameters ($\gamma = 0$). This setup represent the na{\"i}ve learning procedure that could explain investors' behaviour showed in literature\cite{Choi2009}.
The full model, with all the three free parameters, has also been tested against the myopic model to assess 
whether some players are better fitted by a more complex version of the framework. 
Finally the model goodness of fit has been evaluated with the adopted action classification (based on risk) against 500 randomly generated stock classifications. This has been done to test the assumption that players internally classify the stock range into discrete degrees of risk. 

\section{Results}
Results for the test of the hypothesis that RL is a component of decision making are shown in Fig. \ref{fig:models_comparisons}. 
The best set of parameters was found according to MLE through gradient descent search. 
The best model MLE has been compared to the random model MLE using the Likelihood Ratio Test \cite{Crandall1997}. 
The random model MLE is easily estimated as:

\begin{equation}\label{eq:random_MLE_estimation}
	P(D|M_{rand})\ =\ log \prod^{N_t}\frac{1}{3} \ = \ \sum^{N_t}log\frac{1}{3}
\end{equation}

where $N_t$ is the number of transactions for each player in the dataset.
As shown in Fig. \ref{fig:models_comparisons} (a) and (b) 15$\%$ of the players in our dataset is better fitted by a myopic RL model as opposed to a random model. 
In Fig. \ref{fig:models_comparisons}  (c) and (d) we report an improvement in the fitting for some players using a full RL model against the myopic (nested) version of the model. This improvement is not reflected in the comparison of the full RL model against the random model, as shown in Fig. \ref{fig:models_comparisons} (e) and (f). 
Most of the players that can be fitted with our models are well represented by a myopic model.
These results follow what found by Choi et al.\cite{Choi2009} and Huang et al. \cite{Huang2012}.
We made the assumption that players, when faced with the choice to trade many stocks (107 for this task), internally model these in discrete groups of risk using readily available information such as stock historical prices and returns, which in turn are used to estimate their volatility ($\beta_F$).
To test whether this assumption holds true for the players in our dataset, we ran the simpler version of our model on the risk-ranked discretisation and on 500 independent and randomly scrambled discretisations. 
The results are shown in Fig. \ref{fig:ranked_v_scrambled} and are generated using Bayesian Information Criterion (BIC) as a measure for comparison of fitness and Binomial Proportion Confidence Interval calculated with Clopper-Pearson method using \texttt{Matlab 8.4.0.150421 (R2014b)} function \texttt{binofit}. The BIC has been used as the Likelihood Ratio Test can only be used to compare nested models, while in this case the comparison is between models with the same number of parameters that are tested on different data arrangements. 
This procedure estimates the probability that the ranked discretisation is better than the 500 scrambled discretisations ($BIC_{ranked} < BIC_s$, where $BIC_s$ is the BIC for the s-th scrambled).
The results shown in Fig. \ref{fig:ranked_v_scrambled} are for 99$\%$ confidence interval.
The results shown are only for those 7 players who are fitted significantly by the myopic model.
As shown in Fig. \ref{fig:ranked_v_scrambled}, all the players are well above the chance threshold. 
This indicates that risk based on historical data could be considered a proxy for action selection for the players who are well fitted by our RL myopic model.

\section{Conclusion}
We investigated a publicly available dataset consisting of trading transactions operated by players of an investment game. 
We based the discretisation of the actions on the assumption that risk can capture the internal modelling that players operate when facing this task. 
This assumption was shown to hold true and be statistically significant for a subset of the players, 31 out of 46 and specifically for the 7 players who are best fitted by a RL model. 
This could signify that the remaining players might use other types of discretisation techniques based on different measures (or a combination of them) or they do not use technical analysis but fundamental analysis (e.g. using financial statements and reports). 
In this work, we investigated a model which combines two versions of a Reinforcement Learning framework using Q-learning as an update rule and Soft-Max as action-selection policy on a discretised action space according to the risk measure $\beta_F$. 
It is possible that different model combinations, which use different learning rules or different measures of risk, fit the players population in our dataset better. It is also likely that, by restricting our focus on the sell model, we missed some features of what constitutes the reward signal that players receive. In the full version of the game, in fact, players might try to maximise both holdings and cash simultaneously, in order to compete in the ranking. 

The myopic model is a nested version with only two free parameters, representing the learning rate ($\alpha$) and the degree of greediness ($\beta$). 
The full version extends the simpler model with a discount factor ($\gamma$) which regulates how much of the future rewards is taken into account when updating the values of present state-action pairs. 
15$\%$ of the players are well fitted by a RL model with $\gamma = 0$ and there is no significant improvement of fitting by extending this model including gamma as a free parameter.
Previous literature pointed in the direction of investors being na{\"i}ve (short-sighted) \cite{Choi2009, Huang2012} and these results, albeit for a subset of the dataset, confirm this indication.
The hypothesis that RL is a component of the decision making process for some investors is not confirmed as either version of the tested model (short or far-sighted) is statistically better than chance only for a subset of the players. This subset, within this population, is not large enough to draw a statistically meaningful positive result. By means of a Binomial Proportion Confidence Interval calculated with Clopper-Pearson method we get a negative result for the entire population within a 99$\%$ confidence interval (Fig. \ref{fig:population_statistic} in the Appendix).
While this exploratory study gives some perspectives on how Reinforcement Learning can be used to model learning and action-selection for investing problems, future work will focus on different models and risk classification techniques as well as on a deeper investigation of the typical parameters of the best performing players and the correlation of different strategies and performance of stock trading together with a study of different RL models.

\section*{Acknowledgment}

The authors would like to thank their colleagues of the Sheffield Neuroeconomics interdisciplinary research group for insightful discussion.



\bibliographystyle{IEEEtran}
%



\newpage
\appendix
\section{Appendix} \label{App:Appendix}
This manuscript is a correction to the article ``Modelling Stock-market Investors as Reinforcement Learning Agents'' by the same authors, issued in the proceedings of the 2015 IEEE Conference on Evolving and Adaptive Intelligent Systems. 
The corrections include fixing a bug in the script which estimated the probabilities used in the calculation of model fitness. 
In the previous work we applied some constraints and used a different measure:
the number of transactions considered for each player was capped at 25 and the measure of risk used to rank the stocks and classify them into discrete categories was defined by the authors as:

\begin{equation}\label{eq:riskiness_equation}
	R(s_{j}) = \left\lvert {\beta_F}(s_{j}) \cdot \frac{\sigma(s_{j})}{\max \sigma(s)} \right\rvert
\end{equation}

where $\beta_F$ is the financial modelling measure of volatility of a security used in the present work, $\sigma(s_{j})$ is the standard deviation of the j-th stock and the $\max \sigma(s)$ is the largest standard deviation in the stock pool.
This measure of riskiness was used as it was believed to take into account the graphical interpretation of the fluctuations in time series ($\sigma$) and the overall trend of the security compared to the market ($\beta_F$).

\begin{figure}[ht]
\centering
\includegraphics[width=3.5in]{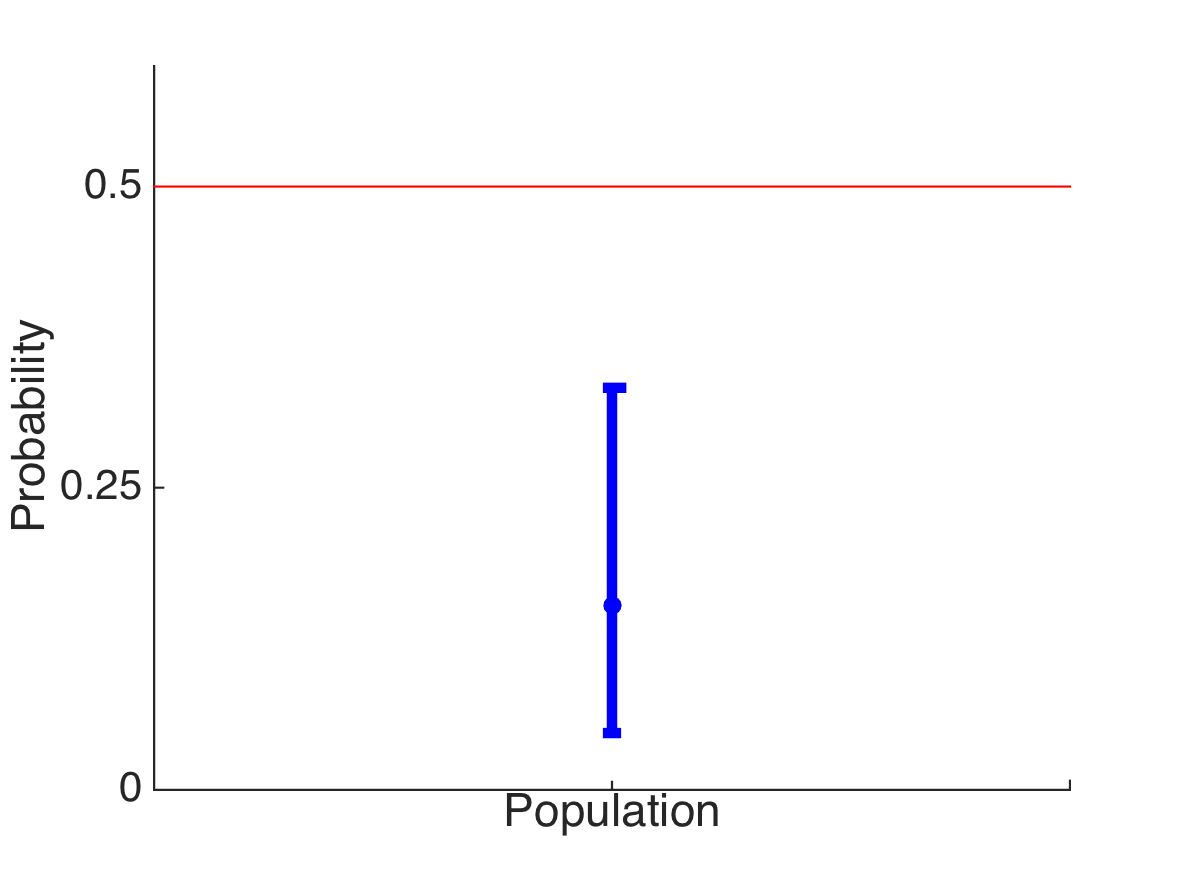}
\caption{Population model fitting statistic using Binomial Proportion Confidence Interval calculated with Clopper-Pearson method. 
The results are negative as only 7 out of 46 players are better than random.
The errorbar represent the 99$\%$ confidence interval for the myopic model to be correctly fitting the players in the dataset. As the errorbar lies entirely below chance threshold we conclude that the models investigated do not correctly represent the behaviour of the players in the data.}
\label{fig:population_statistic}
\end{figure}
\newpage

\begin{figure}[hb]
\centering
\includegraphics[width=3.5in]{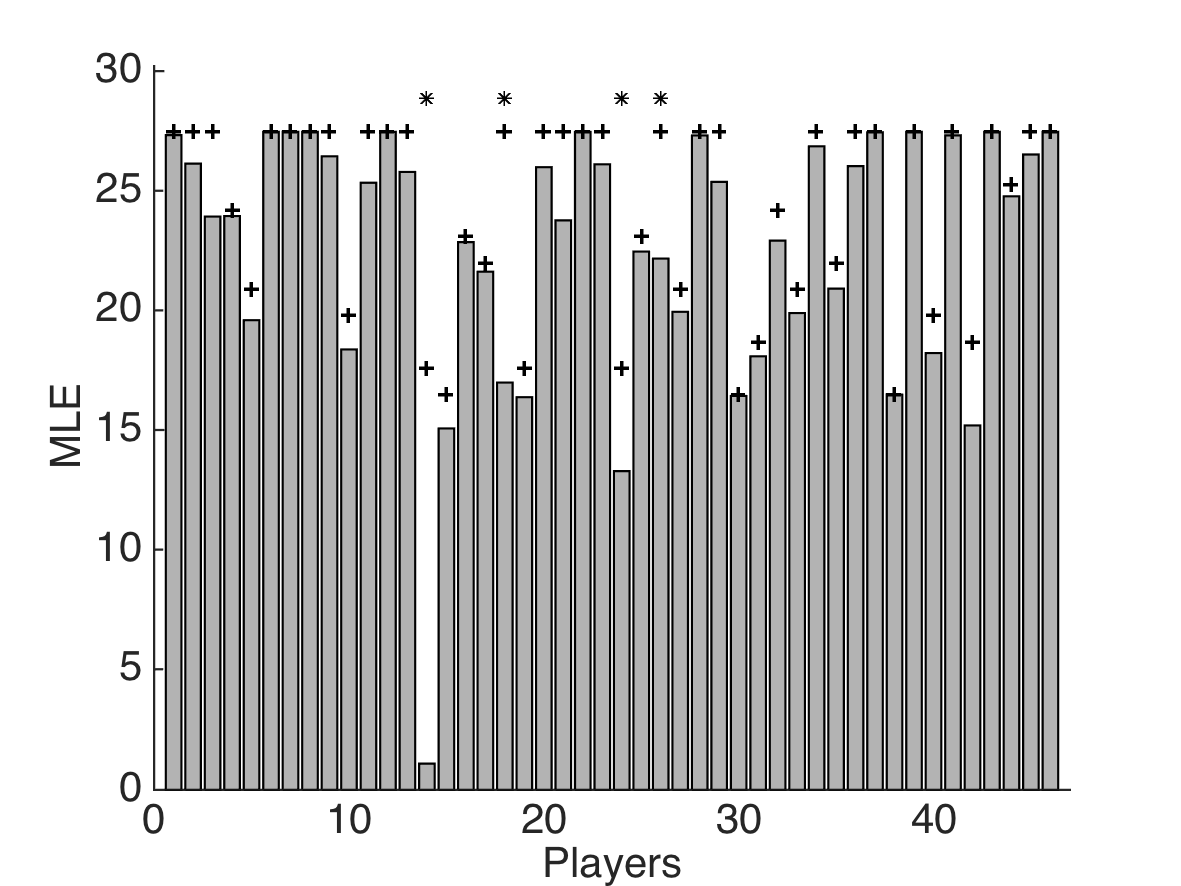}
\caption{Updated model comparison of Fig 1 (a) and (b) in the original paper. This figure shows the comparison of the full RL model (grey bars) against the random model (+ signs) according to their MLE. This is the same comparison as in \ref{fig:models_comparisons} (c) but for the original paper configuration: players transactions capped at 25 and the combination of $\beta_F$ and $\sigma$, as in Eq. \ref{eq:riskiness_equation}, used to determine stocks risk degree and their classification. In this case the portion of players captured by the full RL model is only 8$\%$ (4 players, 14, 18, 24, 26). Only 2 players are captured by the myopic model (14 and 18).}
\label{fig:RL_v_random_original_updated}
\end{figure}

\begin{figure}[ht]
\centering
\includegraphics[width=3.5in]{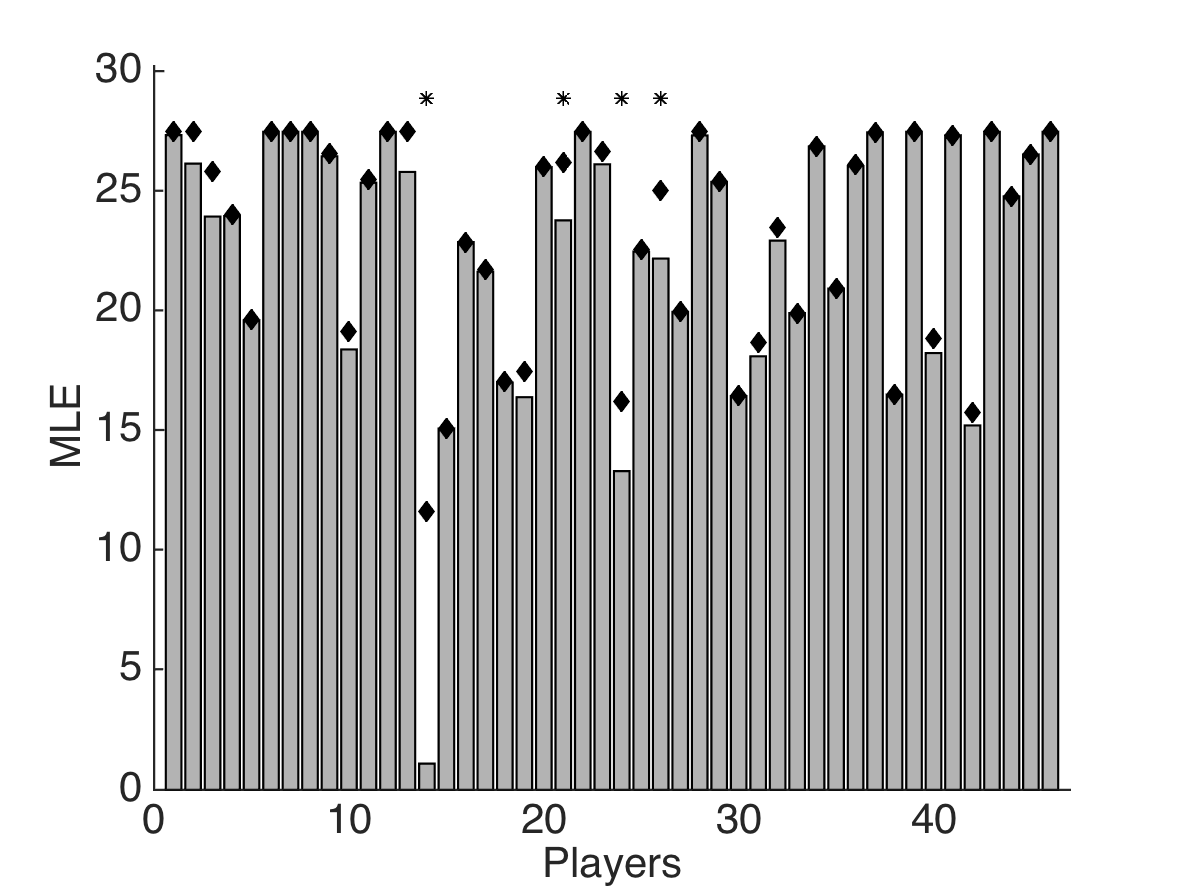}
\caption{Updated model comparison of Fig 1 (c) and (d) in the original paper. This figure shows the comparison of the full RL model (grey bars) against the myopic model (filled diamonds signs) according to their MLE. This is the same comparison as in \ref{fig:models_comparisons} (b) but for the original paper configuration: players transactions capped at 25 and the combination of $\beta_F$ and $\sigma$, as in Eq. \ref{eq:riskiness_equation}, used to determine stocks risk degree and their classification. The 4 players who are significantly better fitted by a full model are 14, 21, 24, 26.}
\label{fig:RL_v_nogamma_original_updated}
\end{figure}

\end{document}